\documentstyle[bbox,epsf]{aipproc}

\def\Journal#1#2#3#4{{#1} {\bf #2}, #3 (#4)}

\def\NPA{{\em Nucl. Phys.} A}
\def\NPB{{\em Nucl. Phys.} B}

\def\PRD{{\em Phys. Rev.} D}

\def\bq{{\bbox{q}}}
\def\bp{{\bbox{p}}}
\def\bK{{\bbox{K}}}
\def\br{{\bbox{r}}}
\def\bx{{\bbox{x}}}

\def\be{\begin{equation}}
\def\ee{\end{equation}}
\def\bea{\begin{eqnarray}}
\def\eea{\end{eqnarray}}

\begin{document}

\title{Testing the Space-Time Structure of Event Generators}

\author{\bf In memory of Klaus Kinder-Geiger}

\author{U. Heinz$^{a,b}$ and U.A. Wiedemann$^{c}$}

\address{$^a$Theoretical Physics Division, CERN, CH-1211 Geneva 23\\
$^b$Institut f\"ur Theoretische Physik, Universit\"at Regensburg,
D-93040 Regensburg\\
$^c$Physics Department, Columbia University, New York, NY 10027}

\maketitle

\vspace*{0.5truecm}

\begin{center}
{\it Combined presentation of the talks given by the two authors at the 
workshop\\
``RHIC Physics and Beyond -- Kay Kay Gee Day''\\
Brookhaven National Laboratory, 23 October 1998}

\vspace*{0.5truecm}

% CERN-TH/99-55~~~~~~~~~~CU-TP-???~~~~~~~~~~hep-ph/9903???
\end{center}

\vspace*{-0.8truecm}

\begin{abstract} 
We report on work done in collaboration with Klaus Kinder-Geiger
and John Ellis which aims at connecting the space-time structure
of event generator simulations with observable output.
\end{abstract}

{\sf \noindent
Klaus Geiger was an important driving force in our search for
understanding the dynamics of ultrarelativistic heavy-ion 
collisions and quark-gluon plasma. An unconventional and unique 
character -- this is how we will remember him. Physics was his 
passion, but he also loved his Porsche and his MG (=``Ma Geiger'').
It is hard for us to accept that by his untimely death 
Klaus, whose research was dedicated so much to the future,
in particular to the RHIC program at Brookhaven, should be
imprisoned forever in the past. As his friends and collaborators,
we will try to carry on his legacy.}

%%%%%%%%%%%%%%%%%%%%%%%%%%%%%%%%%%%%%%%%%%%%%%%%%%%%%%%%%%%%%%
\section{Introduction}
\label{sec1}
%%%%%%%%%%%%%%%%%%%%%%%%%%%%%%%%%%%%%%%%%%%%%%%%%%%%%%%%%%%%%%

The main focus of Klaus'\ work during the last years of his life 
was the event generator VNI which describes relativistic 
heavy-ion collisions and high energy particle collisions in terms 
of a perturbative parton shower Monte Carlo in phase-space combined 
with a simple space-time hadronization prescription. A partonic 
starting point is certainly required at ultrarelativistic collision 
energies, $\sqrt{s} \gtrsim 100\,A$ GeV, where perturbative contributions 
start to account for a significant part of the measured particle 
production. With his Parton Cascade Model \cite{PCM}, later amended
by a space-time hadronization algorithm \cite{EG95} and made publicly 
available in form of the event generator VNI \cite{VNI}, Klaus Geiger 
led the way. 

From summer 1997 until Klaus'\ death we collaborated with him and John 
Ellis on implementing an afterburner for Bose-Einstein correlations into
this code and testing it in simulations for $e^+e^-$ collisions at LEP\,I
and LEP\,II collisions. Our goal was to prepare for an experimental test
of the space-time dynamics predicted by VNI and to provide theoretical 
guidance for multi-dimensional Bose-Einstein analyses of various event 
classes with hadronic final states generated at LEP. This work 
\cite{WEHG98,GEHW98} was interrupted prematurely; our contribution
gives an account of the present status.

VNI distinguishes itself from many other high energy event Monte Carlos
by following the event history in phase-space, not only in momentum 
space. {\em Space-time aspects} enter in the numerical simulation in 
various ways: On the {\it microscopic} level, the rescattering between
produced partons or hadrons is controlled by geometric cross sections, 
and hadronization is modelled geometrically by requiring partons to 
get (in their pair rest frame) closer than 0.8 fm in coordinate space 
in order to form a hadronic cluster. On a {\it macroscopic} level
this leads to a strong density dependence of particle production and 
absorption rates, further affected by the collective expansion of the
system which is generated by the rescattering. Finally, quantum 
statistical effects among the produced final state particles, in 
particular Bose-Einstein correlations in momentum space between pairs 
of identical pions or kaons, depend on the {\em phase-space density} of 
the system at the point of decoupling and thus on both the 
momentum-space and space-time structure of the event at ``freeze-out'' 
(i.e. at the point of the last strong interaction between the particles).

In this sense all measured quantities in a heavy-ion collision depend
to some extent on the space-time structure of the reaction zone. The 
crucial question is, however, whether they are sufficiently sensitive
to such aspects to allow for a reconstruction of the space-time 
geometry and dynamics. The successful reproduction of single-particle
yields and spectra by event generators with different space-time
features, or none at all like the popular JETSET \cite{LS95} and 
PYTHIA \cite{PYTHIA} generators, seems to argue against such a 
possibility. Two-particle correlations in momentum space, however,
are sensitive to both the geometric extension of the ``fireball'' at 
freeze-out and to its collective expansion dynamics \cite{WH99}. 
The latter affects the two-particle spectra via so-called 
``$x$-$p$-correlations'' in the emission function $S(x,p)$. This 
function is the quantum mechanical analogue of the single-particle 
phase-space density of the source at freeze-out: collective expansion 
correlates the average direction and magnitude of the momenta of the emitted 
particles with their emission points. A detailed analysis of 2-pion and
2-kaon correlations in relativistic heavy-ion collisions at the Brookhaven
AGS and the CERN SPS has recently led to an unambiguous demonstration of 
strong collective dynamics of the fireballs created in these experiments
\cite{HJ99}. On a finer level, however, there remain a number of 
open physical questions whose resolution requires two-particle correlation 
data of similar quality and detail from elementary particle collisions.
Such data do not exist, and another motivation for our work with Klaus 
Geiger was therefore to provide stimulation for similar experimental 
analyses of high-statistics high energy data samples like the $\sim 20$ 
million hadronic $Z^0$ decays collected at LEP\,I.

For sufficiently high secondary particle multiplicities (i.e. at very 
high energies or for large collision systems) it is reasonable to assume 
that the two particles of a selected pair were emitted independently. 
This allows to express the two-particle correlation function in terms
of the single-particle Wigner density $S(x,p)$ of the source \cite{WH99}.
Neglecting final state interactions (assuming that they can be corrected 
for experimentally \cite{WH99,HJ99} or theoretically \cite{AHR97} at a 
later stage), it is given by \cite{WH99}
  \begin{equation}
     C(\bq,\bK) = {\cal N} \left(1 + 
     {\left\vert \int d^4x\, S(x,K)\, e^{iq{\cdot}x}\right\vert^2 
      \over
      \int d^4x\, S(x,p_1)\, \,\int d^4y\, S(y,p_2)}\right)\, .
   \label{eq3}
  \end{equation}
Here $p_1,p_2$ are the on-shell momenta of the two particles in the pair
while $K=(p_1+p_2)/2$ and $q=p_1-p_2$ are their average and relative 
4-momenta. The numerator in the second term stands for the product
of the measured single-particle spectra
  \begin{equation}
     E_p {dN\over d^3p} = \int d^4x\, S(x,p)
  \label{eq3a}
  \end{equation}
where $p^0=E_p=\sqrt{m^2+\bp^2}$. The normalization ${\cal N}$ will 
be discussed below. 

If, for a given momentum $K$, the space-time dependence of $S(x,K)$ can 
be characterized with reasonable accuracy by a single set of rms widths
(i.e. the particle emission is not characterized by several widely
differing length scales), one can approximate $S(x,K)$ by a Gaussian in 
$x$. The correlation function then takes the simple form \cite{WH99}
  \begin{equation}
     C(\bq,\bK) = {\cal N} \left(1 + 
     e^{- q^\mu q^\nu \langle \tilde x_\mu \tilde x_\nu\rangle(K)}
     \right)\, ,
   \label{eq3b}
  \end{equation}
where $\langle \tilde x_\mu \tilde x_\nu \rangle(K)$ describes the
second space-time moments (rms widths or ``homogeneity lengths'') of 
the {\em effective source} of particles with momentum $K$. Since the 
two measured particles are on-shell, $p_1^2=p_2^2=m^2$, $K\cdot q=0$ 
and only 3 of the 4 components of $q$ are independent. Different
choices for the independent components lead to different Gaussian 
parametrizations of the correlation function \cite{WH99}. We will here 
use the Cartesian parametrization which eliminates 
$q^0=\bbox{\beta}\cdot\bq$ where $\bbox{\beta}= \bK/K^0$ is 
(approximately) the velocity of the particle pair. $\bq$ is
decomposed into its Cartesian components $(q_o,q_s,q_l)$ where
$l$ denotes the ``longitudinal'' direction (in heavy-ion collisions this 
is the direction of the beam axis, in $e^+e^-$ collisions the direction
of the thrust axis), $o$ denotes the outward direction, fixed by the
azimuthal orientation of the transverse pair momentum $\bK_\perp$
around the $l$-axis, and $s$ denotes the third Cartesian (sideward) 
direction (defined by $K_s=0$, $\bK=(K_\perp,0,K_l)$).

Eliminating $q^0$ from the exponent in (\ref{eq3b}) leaves a sum over 6
terms; using furthermore the azimuthal symmetry of the event sample 
around the $l$-axis allows to further reduce this to 4 terms involving
certain algebraic sums \cite{WH99} of the rms widths 
$\langle \tilde x_\mu \tilde x_\nu \rangle(K)$. So far, however,
Bose-Einstein correlations in elementary particle collisions 
have been parametrized by much simpler Gaussian forms, involving only 
one or at most two ($K$-independent!) size parameters. Such incomplete
parametrizations have the fundamental disadvantage that the fit 
parameter(s) mix(es) the interesting space-time information contained
in the rms widths $\langle \tilde x_\mu \tilde x_\nu \rangle(K)$ of
the effective source of particles with momentum $K$ in such a way that 
they can no longer be recovered \cite{He96}. An important goal of our 
work with Klaus was to make predictions for the multidimensional shape
of the correlation function $C(\bq,\bK)$, by calculating the complete 
set of size parameters and their $K$-dependence. 

The latter is particularly interesting since it signals 
$x$-$K$-correlations in the emission function $S(x,K)$. While in 
heavy-ion collisions the dominant mechanism for such correlations 
appears to be collective expansion of the fireball \cite{He96,WH99,HJ99},
contaminations from resonance decays after freeze-out \cite{resonances}
and from $x$-$p$-correlations in the primary hadron formation process 
\cite{BZ99} (e.g. from string fragmentation) are known to exist. The 
latter are expected to play a much bigger role in elementary particle 
collisions where a multidimensional analysis of Bose-Einstein 
correlations may help to isolate them. This would provide crucial input 
into a quantitative discussion of Bose-Einstein correlations from 
heavy-ion collisions where, contrary to elementary particle collisions 
where the multiplicities are much lower, most resonance decays can 
{\em not} be reconstructed experimentally and their effects must 
therefore be simulated.  

%%%%%%%%%%%%%%%%%%%%%%%%%%%%%%%%%%%%%%%%%%%%%%%%%%%%%%%%%%%%%%%%%%%
\section{From phase-space densities to momentum correlations}
\label{sec2}
%%%%%%%%%%%%%%%%%%%%%%%%%%%%%%%%%%%%%%%%%%%%%%%%%%%%%%%%%%%%%%%%%%%

Event generators like VNI evolve classical {\em probabilities}, not 
quantum mechanical {\em amplitudes} calculated from properly symmetrized 
many-particle wave functions. The Bose-Einstein correlations among 
pairs of identical pions or kaons must thus be included 
{\em a posteriori}. For notational convenience, we restrict the 
following discussion to a single particle species, say like-sign 
pions. Let us assume that we have generated $N_{\rm ev}$ collision 
events, and let the $m$th event ($1\leq m \leq N_{\rm ev}$) consist 
of $N_m$ such pions in the final state, emitted as free particles 
from the phase-space points $\lbrace ({\check{\br}}_i,{\check{\bp}}_i, 
 {\check{t}}_i) \rbrace_{i\in [1,N_m]}$. For $N_{\rm ev}$ events, 
the event generator thus simulates a classical phase-space distribution
 \begin{equation}
 \label{eq2}
   \rho_{\rm class}(\br,\bp,t) =
   {1\over N_{\rm ev}} \sum_{m=1}^{N_{\rm ev}} \sum_{i=1}^{N_m}
   \delta^{(3)}(\br-{\check{\br}}_i)\,
   \delta^{(3)}(\bp-{\check{\bp}}_i)\, 
   \delta(t-{\check{t}}_i)\, .
 \end{equation}
What is needed to calculate two-particle Bose-Einstein correlations
according to (\ref{eq3}) is a prescription which relates this classical 
phase-space density with the quantum mechanical single-particle Wigner 
density $S(x,p)$ of the pion source which is supposed to be Monte Carlo 
simulated by the event generator. In sections~\ref{sec2a} and~\ref{sec2b}, 
we focus on two different interpretations of the event generator output
which we call ``classical'' and ``quantum'', respectively, although these 
names should not distract from the fact that conceptually both are on an 
equal footing. Each one of them leads to a different algorithm for 
calculating the two-particle correlation function from the event 
generator output:
  \begin{equation}
    \Bigl\lbrace
    \lbrace (\check{\bf r}_i, \check{\bf p}_i, 
    \check{t}_i)\rbrace_{i\in [1,N_m]}
    \Bigr\rbrace_{m\in [1,N_{\rm ev}]}
    \Longrightarrow \ C(\bq,\bK)\, ,
    \label{eq1}
  \end{equation}
Before discussing them we list three requirements which any such
algorithm should fulfill:
  \begin{enumerate}
    \item
    Since the event generator evolves classical probabilities,
    not symmetrized production amplitudes, the generated momenta 
    $\check{\bp}_i$, $\check{\bp}_j$ do not show the enhanced 
    probability at low relative momenta $\check{\bp}_i-\check{\bp}_j$
    characteristic for Bose-Einstein final state symmetrization. The
    prescriptions (\ref{eq1}) should calculate this effect a 
    posteriori.
    \item
    The strength of Bose-Einstein correlations depends on the
    distance of the identical particles in phase-space, not
    in momentum space. We thus require the prescriptions
    (\ref{eq1}) to use the entire phase-space information,
    and not only the generated momentum information. `Weighting' 
    or `shifting' prescriptions which are based only on the
    latter \cite{LS95} may successfully match the measured 
    momentum correlations but obviously do not allow to test the
    simulated space-time structure.
    \item
    In general, Bose-Einstein statistics can affect particle
    multiplicity distributions during the particle production
    process but classical event generators do not include such
    effects explicitly. Nevertheless they are tuned to reproduce
    the measured average particle multiplicities $\langle N \rangle$. 
    In order not to destroy this tuning we require that the 
    prescription (\ref{eq1}) conserves the single-particle 
    multiplicities. If the event generator were also tuned to reproduce
    the average multiplicity of identical particle {\em pairs}, 
    $\langle N(N-1)/2\rangle$, i.e. to reproduce not only the mean, but
    also the width of the multiplicity distribution, then the
    prescription (\ref{eq1}) should not change that either. Interpreting 
    the correlator as a factor which relates the measured two-particle
    differential cross section to the one simulated by the
    event generator, $d\sigma_{\rm meas}/d^3p_1\, d^3p_2
    = C(\bq,\bK)\, d\sigma_{\rm sim}/d^3p_1\, d^3p_2$, 
    this then implies~\cite{WEHG98,LS95,W98} that ${\cal N} < 1$ in 
    (\ref{eq3}). The algorithms discussed below do not satisfy this
    last requirement, i.e. in general they change the width of the 
    multiplicity distribution. However, since the space-time analysis of 
    correlation data can be based entirely on the momentum dependence of 
    $C(\bq,\bK)$, irrespective of its absolute normalization, this
    does not matter for our purposes. 
  \end{enumerate}
%
%%%%%%%%%%%%%%%%%%%%%%%%%%%%%%%%%%%%%%%%%%%%%%%%%%%%%%%%%%%%%%%%%%%%%
\subsection{``Classical'' interpretation of event generator output}
\label{sec2b}
%%%%%%%%%%%%%%%%%%%%%%%%%%%%%%%%%%%%%%%%%%%%%%%%%%%%%%%%%%%%%%%%%%%%%
%
In the ``classical'' interpretation~\cite{ZWSH97,GEHW98}, 
the phase-space points $(\check{\br}_i, \check{\bp}_i, \check{t}_i)$ 
are seen as a discrete approximation of the on-shell Wigner 
phase-space density $S(x,p)$: 
  \begin{equation}
    S(x,\bp) = \rho_{\rm class}(\bx,\bp,t)\, .
    \label{eq10}
  \end{equation}
This defines both the two-particle correlator via (\ref{eq3}) and the
one-particle spectrum via (\ref{eq3a}). For a numerical implementation 
one must replace the momentum-space $\delta$-functions in (\ref{eq2})
by normalized ``bin functions'' with finite width $\epsilon$ of 
rectangular \cite{ZWSH97} or Gaussian~\cite{GEHW98} shape, e.g.
  \begin{equation}
    \delta^{(\epsilon)}_{\check{\bp}_i,{\bp}}
    = {1\over (\pi\epsilon^2)^{3/2}}
      \exp \left( - (\check{\bp}_i - \bp)^2/\epsilon^2\right)\, ,
    \label{eq11}
  \end{equation}
which reduce to the original $\delta$-function in the limit 
$\epsilon \to 0$. The one-particle spectrum and two-particle correlator 
then read \cite{ZWSH97,GEHW98}
  \begin{equation}
     E_p {dN\over d^3p} = \int d^4x\, S(x,{\bp}) =
       {E_p \over N_{\rm ev}} \sum_{m=1}^{N_{\rm ev}}
       \sum_{i=1}^{N_m} \delta^{(\epsilon)}_{\check{\bp}_i,{\bp}}\, ,
   \label{eq12} 
  \end{equation}
  \begin{equation}
   C(\bq,\bK) = 1 + { \sum_{m=1}^{N_{\rm ev}} \left[
   \left\vert \sum_{i=1}^{N_m} 
       \delta^{(\epsilon)}_{\check{\bp}_i,{\bK}}\, 
       e^{i(q^0\check{t}_i - \bq{\cdot}\check\br_i)} \right\vert^2 
        - \sum_{i=1}^{N_m} \left(\delta^{(\epsilon)}_{\check{\bp}_i,\bK}
              \right)^2
   \right]
   \over
   \sum_{m=1}^{N_{\rm ev}} \left[ \left(\sum_{i=1}^{N_m} 
       \delta^{(\epsilon)}_{\check{\bp}_i,\bp_1}\right)
       \left(\sum_{j=1}^{N_m} 
         \delta^{(\epsilon)}_{\check{\bp}_j,\bp_2}\right)
   - \sum_{i=1}^{N_m} \delta^{(\epsilon)}_{\check{\bp}_i,\bp_1}
                      \delta^{(\epsilon)}_{\check{\bp}_i,\bp_2}
   \right]}\, .
 \label{eq13}
 \end{equation}
The correlator (\ref{eq13}) is the discretized version of the Fourier 
integral in (\ref{eq3}). The subtracted terms in the numerator and 
denominator remove discretization errors which would amount to pairs 
constructed from the same particles. This ``classical'' algorithm, as well
as the ``quantum'' version discussed below, is numerically efficient 
since it involves only $O(N_m)$ manipulations per event. The 
calculated observables, while being discrete functions of the input, 
are continuous functions of the measured momenta, i.e. no binning 
in $\bq,\bK$ is necessary.
%%%%%%%%%%%%%%%%%%%%%%%%%%%%%%%%%%%%%%%%%%%%%%%%%%%%%%%%%%%%%%%%%%%%%
\subsection{``Quantum'' interpretation of event generator output}
\label{sec2a}
%%%%%%%%%%%%%%%%%%%%%%%%%%%%%%%%%%%%%%%%%%%%%%%%%%%%%%%%%%%%%%%%%%%%%

The ``quantum'' interpretation \cite{ZWSH97,Weal97,W98,GEHW98} of
the event generator output starts from the observation that, for a given 
event, i.e. a single term in the sum of (\ref{eq2}), the simultaneous
sharp definition of the particle momenta and positions at emission 
violates the uncertainty relation. In the limit $N_{\rm ev}\to\infty$
the sum is still expected to be a smooth phase-space function and, to 
the extent that the event generator provides an accurate simulation
of the underlying QCD quantum dynamics, it is also expected to respect
the uncertainty constraints on any allowed Wigner density. Since in 
practice, however, one has to work with finite numbers of events, one 
may wish to ensure consistency with the uncertainty principle on the 
single-event level.  
 
This is achieved \cite{Weal97,ZWSH97,MP97} by associating the centers 
of single-particle wave packets with the set of generated phase-space 
points $\lbrace \lbrace (\check{\br}_i, \check{\bp}_i, \check{t}_i)
\rbrace_{i\in [1,N_m]} \rbrace_{m\in [1,N_{\rm ev}]}$:
  \begin{eqnarray}
    (\check{\bp}_i, \check{\br}_i, \check{t}_i) \longrightarrow
    f_i({\bx},\check{t}_i) 
    = {1\over (\pi\sigma^2)^{\footnotesize{3/4}} }
    e^{-{1\over 2\sigma^2}\, 
      (\bx-\check{\br}_i)^2 + i\check{\bp}_i\cdot \bx }\, .
    \label{eq4}
  \end{eqnarray}
The $f_i$ describe quantum mechanically best-localized states, i.e. 
they saturate the Heisenberg uncertainty relation with 
$\Delta x_i = \sigma/\sqrt{2}$ and $\Delta p_i = 1/\sqrt{2}\sigma$.

Taking only two-particle symmetrized contributions into account
(``pair approximation'' \cite{W98}), all spectra can be written 
\cite{ZWSH97,Weal97} in terms of the single-particle spectrum 
$s_i(\bp)$ corresponding to an individual wave packet at phase-space 
point $i$:
 \begin{eqnarray}
   E_p {dN\over d^3p} &=& {E_p \over N_{\rm ev}} \sum_{m=1}^{N_{\rm ev}}
   \nu_m(\bp) = {E_p \over N_{\rm ev}} 
   \sum_{m=1}^{N_{\rm ev}} \sum_{i=1}^{N_m} s_i(\bp)\, ,
 \label{eq5} \\
    C(\bq,\bK) &=&  1 + e^{-{1\over2}\sigma^2 \bq^2} 
     { \sum_{m=1}^{N_{\rm ev}} \left[ 
       \left\vert \sum_{i=1}^{N_m} s_i(\bK)\,
         e^{i(q^0\check{t}_i - \bq\cdot\check{\br}_i)} \right\vert^2
      - \sum_{i=1}^{N_m} s_i^2(\bK) \right]
      \over 
        \sum_{m=1}^{N_{\rm ev}} \left[ \nu_m(\bp_1)\, \nu_m(\bp_2) 
      - \sum_{i=1}^{N_m} s_i(\bp_1)\,s_i(\bp_2) \right]}\, ,
 \label{eq6} \\
   s_i(\bp) &=& \left(\sigma^2 \over\pi\right)^{3/2}\, 
   e^{ - \sigma^2 (\bp-\check{\bp}_i)^2}\, .
 \label{eq7}
 \end{eqnarray}
Again, the subtracted terms in the numerator and denominator of 
$C(\bq,\bK)$ are finite multiplicity corrections which become 
negligible for large particle multiplicities \cite{Weal97}.
This algorithm is consistent with an emission function $S(x,\bK)$ 
which is obtained by folding the classical distribution 
$\rho_{\rm class}$ of wave packet centers with the 
Wigner density $s_0(x,\bK)$ of a single wave packet: 
  \begin{eqnarray}
    S(x,\bK) &=& \int d^3\check{r}_i\, d^3\check{p}_i\, d\check{t_i}\,
    \rho_{\rm class}(\check{\br}_i,\check{\bp}_i,\check{t_i})\, 
    s_0(\bx-\check{\br}_i,t-\check{t_i},\bK-\check{\bp}_i)\, ,
  \label{eq8}\\
    s_0(x,\bK) &=&
    {1\over \pi^3}\, \delta(t)\,
    e^{-{1\over \sigma^2}\bx^2 - \sigma^2 \bK^2}\, .
  \label{eq9}
  \end{eqnarray}
The latter saturates the uncertainty relation with a spatial uncertainty 
$\sigma$ and a momentum uncertainty $1/\sigma$, and the folding ensures
that now $S(x,\bK)$ is always quantum mechanically consistent. However, 
in this algorithm both the one- and two-particle spectra depend on the 
wave packet width $\sigma$. The role of this parameter will be discussed 
in the context of our toy model study in section~\ref{sec3}.

The ``classical'' and ``quantum'' algorithms differ only with respect to 
two points:
  \begin{enumerate}
    \item
      The ``classical'' algorithm has no analogue for the
      Gaussian prefactor $\exp\left(-{1\over 2}\sigma^2\,\bq^2/\right)$
      in (\ref{eq6}) which is a genuine quantum effect stemming from
      the quantum mechanical localization properties of (\ref{eq4}).
    \item
      The Gaussian single-particle distributions $s_i(\bp)$ in
      the ``quantum'' algorithm are the quantum analogues of the
      ``bin functions'' in the ``classical'' agorithm. With the Gaussian 
      bin functions (\ref{eq11}) the two agree for the choice 
      $\sigma = 1/\epsilon$. Finite event statistics puts a lower 
      practical limit on $\epsilon$ in the ``classical'' algorithm, but
      to get accurate spectra one should try to choose $\epsilon$ as 
      small as possible, by simulating sufficiently many events. In 
      contrast, $\sigma$ in the ``quantum'' algorithm is the finite
      spatial width of the single-particle wave packets, and the limit 
      $\sigma \to \infty$ (which corresponds to $\epsilon \to 0$) is not 
      physically relevant: according to (\ref{eq8}) it amounts to an 
      emission function with infinite spatial extension and thus 
      to a correlator \cite{Weal97}
      $\lim_{\sigma\to\infty} C(\bq,\bK) = 1 + \delta_{\bq,\bbox{0}}$. 
  \end{enumerate}

%%%%%%%%%%%%%%%%%%%%%%%%%%%%%%%%%%%%%%%%%%%%%%%%%%%%%%%%%%%%%%%%%%
\section{The Zajc Model}
\label{sec3}
%%%%%%%%%%%%%%%%%%%%%%%%%%%%%%%%%%%%%%%%%%%%%%%%%%%%%%%%%%%%%%%%%%

Before describing realistic event generator simulations we 
discuss some analytical results for a classical toy model 
distribution $\rho_{\rm class}$. These are then used to test
the algorithms of sections~\ref{sec2a} and ~\ref{sec2b},
by applying those to sets of phase-space points
$\lbrace (\check{\bf r}_i,\check{\bf p}_i,\check{t}_i)\rbrace_{i\in[1,N_m]}$ 
generated from the model distribution with a Monte Carlo procedure.
In this way we can make quantitative statements about i) the 
$\sigma$-dependence of the ``quantum'' algorithm, ii) the 
$\epsilon$-dependence of the ``classical'' algorithm (especially: how small 
$\epsilon$ has to be chosen to extract the limit $\epsilon \to 0$ 
numerically) and iii) the statistical requirements for the algorithms 
to work.

The toy model, first introduced by Zajc~\cite{Z93}, reads:
  \begin{eqnarray}
    \rho_{\rm class}^{\rm Zajc}(\br,\bp,t) &=& 
    {\cal N}_s\, \delta(t)\,\exp \left\{ -\frac{1}{2(1-s^2)} 
             \left(    \frac{\br^2}{R_0^2} 
                 -2s\frac{\br\cdot\bp}{R_0 P_0}
                    +  \frac{\bp^2}{P_0^2}
                 \right) \right\}  ,
  \label{eq14} \\
    {\cal N}_s  &=& E_p\, 
    {N\over (2\pi R_s P_0)^3} \, , \quad R_s = R_0\sqrt{1-s^2}\,. 
    \label{eq15}
  \end{eqnarray}
The distribution is normalized to a total event multiplicity $N$. 
The parameter $s$ smoothly interpolates between completely 
position-momentum correlated and uncorrelated sources. 
For $s \to 0$, the position-momentum correlation vanishes
and we are left with the product of two Gaussians in $\br$ and
$\bp$. In the limit $s\to 1$
  \begin{equation}
  \label{eq16}
    \lim_{s\to 1}  \rho_{\rm class}^{\rm Zajc}({\bf r},{\bf p},t) 
       \sim \delta^{(3)}
    \left(\frac{{\bf r}}{R_0} - \frac{{\bf p}}{P_0}\right)\,\delta(t)\, ,
  \end{equation}
the position-momentum correlation is perfect. The total phase-space 
volume of the distribution $V_{\rm p.s.} = (2R_sP_0)^3$ vanishes
for $s \to 1$. This strong $s$-dependence allows to study the performance 
of our numerical algorithms for different phase space volumes. In the 
following subsections, we discuss the $s$-dependence of the one-particle 
spectrum and two-particle correlator, focussing in sections~\ref{sec3a} 
and \ref{sec3b} on analytical results, and comparing these in 
section~\ref{sec3c} to numerical calculations.

%%%%%%%%%%%%%%%%%%%%%%%%%%%%%%%%%%%%%%%%%%%%%%%%%%%%%%%%%%%%%%%%%
\subsection{The Zajc model in the ``classical'' algorithm}
\label{sec3a}
%%%%%%%%%%%%%%%%%%%%%%%%%%%%%%%%%%%%%%%%%%%%%%%%%%%%%%%%%%%%%%%%%

Inserting (\ref{eq14}) into (\ref{eq10})-(\ref{eq13}) one finds \cite{GEHW98}
  \begin{eqnarray}
     E_p{dN\over d^3p} &=& E_p {N\over (2\pi P_0^2)^{3/2}}\, 
                        \exp\left( -{\bp^2\over 2P_0^2}\right)\, ,
  \label{eq18}\\
     C(\bq,\bK) &=& 1 + \exp\Bigl(-\bq^2 R_{\rm class}^2(\epsilon)\Bigr)\, ,
  \label{eq19}\\
     R_{\rm class}^2(\epsilon) &=& 
     {R_s^2\over 1 + \epsilon^2/(2P_0^2)}
     \left( 1 + {\epsilon^2\over 2P_0^2(1-s^2)}
              -{1\over (2 R_s P_0)^2}\right).
  \label{eq20}
  \end{eqnarray}
We recover the physical HBT radius $R_{\rm class}^2 = 
R_s^2\big(1-1/(2R_sP_0)^2\big)$ from (\ref{eq20}) in the limit 
$\epsilon \to 0$ or by inserting (\ref{eq10}) directly into (\ref{eq3}).
The remarkable fact is that for sufficiently large $s$~\cite{Z93},
  \begin{equation}
    s > s_{\rm crit} = \sqrt{ 1 - {1\over (2R_0P_0)^2}}\, ,
    \label{eq21}
  \end{equation}
the HBT radius $R_{\rm class}^2$ becomes negative and the two-particle
correlator shows an unphysical rise of the correlation function with 
increasing $\bq^2$. The change of sign in (\ref{eq19}) seems to be 
related to the violation of the uncertainty relation by the emission 
function for large $s$ when $\br$ and $\bp$ become strongly correlated.
At the critical value $s_{\rm crit}$ the phase-space volume $V_{\rm p.s.}$
of the source drops below 1. Only for $s < s_{\rm crit}$ the
distribution $\rho_{\rm class}^{\rm Zajc}$ is a quantum mechanically
allowed emission function, i.e. a Wigner density.

The practical importance of the Zajc model in the unphysical limit 
$s \geq s_{\rm crit}$ is that it provides an extreme testing ground for
our numerical algorithms. Analytically, we conclude already from
(\ref{eq20}) that in order to be close to the physical limit 
the bin width $\epsilon$ has to be small on the scale of the 
width of the generated momentum distribution,
  \begin{equation}
    \epsilon \ll \sqrt{2}\, P_0\, .
    \label{eq22}
  \end{equation}
In the above toy model this requirement is independent of the strength 
of position-momentum correlations in the source.

%%%%%%%%%%%%%%%%%%%%%%%%%%%%%%%%%%%%%%%%%%%%%%%%%%%%%%%%%%%%%%%%%
\subsection{The Zajc model in the ``quantum'' algorithm}
\label{sec3b}
%%%%%%%%%%%%%%%%%%%%%%%%%%%%%%%%%%%%%%%%%%%%%%%%%%%%%%%%%%%%%%%%%

Inserting the model distribution $\rho_{\rm class}^{\rm Zajc}$ 
into (\ref{eq8}) (instead of (\ref{eq10})), we find
  \begin{eqnarray}
     E_p{dN\over d^3p} &=& E_p {N\over (2\pi P^2)^{3/2}} 
     \exp\left( -{\bp^2\over 2P^2}\right)\, ,
  \label{eq23}\\
     C(\bq,\bK) &=& 1 + \exp\Big( -\bq^2 R_{\rm qm}^2\Big)\, ,
  \label{eq24} \\
     R_{\rm qm}^2 &=& R^2 \left(1- {1\over (2RP)^2}\right)\, ,
  \label{eq25} \\
     R^2 &=& R_2^2 + {\sigma^2\over 2}\, , \quad 
     P^2 = P_0^2 + {1\over 2\sigma^2}\,.
  \label{eq25a}
  \end{eqnarray}
In this case, $R$ and $P$ satisfy $2RP\geq 1$ independent of
the value of $\sigma$ and, in contrast to the ``classical'' algorithm, 
the radius parameter $R_{\rm qm}^2$ is now always positive, irrespective 
of the value of $s$. Even if the classical distribution 
$\rho_{\rm class}^{\rm Zajc}(\check{\br},\check{\bp},\check{t})$ 
is sharply localized in phase-space, its folding with minimum uncertainty 
wave packets leads to a quantum mechanically allowed emission function 
$S(x,\bp)$ and to a correlator with falls off with increasing
$\bq^2$ as expected. However, the spread of the one-particle
momentum spectrum (\ref{eq23}) receives an additional contribution
$1/\sigma^2$. Choosing $\sigma$ too small increases this term beyond the
phenomenologically reasonable values, choosing it too large widens
the corresponding HBT radius parameters significantly. It was
argued~\cite{Weal97} that $\sigma$ can be interpreted as quantum
mechanical ``size'' of the particle, $\sigma \approx 1$ fm. Given 
the heuristic nature of these arguments and the significant
modifications this implies for the spectra (\ref{eq23}) and
(\ref{eq24}), it is however fair to say that presently $\sigma$
mainly plays the role of a regulator of unwanted violations of the 
quantum mechanical uncertainty principle while a deeper understanding
of its origin in the particle production dynamics is still missing.

%%%%%%%%%%%%%%%%%%%%%%%%%%%%%%%%%%%%%%%%%%%%%%%%%%%%%%%%%%%%%%%%%
\subsection{Numerical simulations in the Zajc model}
\label{sec3c}
%%%%%%%%%%%%%%%%%%%%%%%%%%%%%%%%%%%%%%%%%%%%%%%%%%%%%%%%%%%%%%%%%

We have studied the performance of our Bose-Einstein algorithms
by generating with a random number generator sets
$\lbrace{(\check{\br}_i,\check{\bp}_i,\check{t}_i)}\rbrace_{i\in [1,N_m]}$ 
of phase-space points according to the distribution 
$\rho_{\rm class}^{\rm Zajc}$ and comparing the numerical results 
of our algorithms to the analytical expressions of section~\ref{sec3a} 
and~\ref{sec3b}. 

%%%%%%%%%%%%%%%%%%%%%%%%%%%%%%%%%%%%%%%%%%%%%%%%%%%%%%%%%%%%%%%%%%%%
\begin{figure}[ht]\epsfxsize=9.5cm 
\centerline{\epsfbox{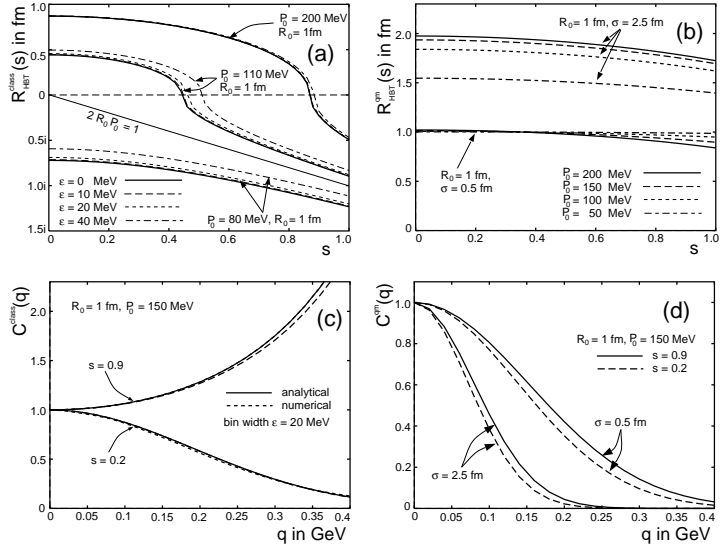}}
\caption{
Generic properties of the one-dimensional Zajc model. (a): HBT-radius 
parameter (\ref{eq20}) of the ``classical'' interpretation as a function 
of the position-momentum correlation $s$. The plot shows the
HBT radius for different combinations of the model parameters
$R_0$ and $P_0$, and for different sizes of the bin width $\epsilon$
used in the numerical implementation. (b): Same as (a) for
the ``quantum'' version (\ref{eq25}) of the model. The dependence
of the HBT radius on the wave packet width $\sigma$ is clearly seen.
(c) and (d): The two-particle correlator in the ``classical'' (c) and
``quantum'' (d) version of the model for different sets
of model parameters. The numerical results are obtained for $50$
events of multiplicity $1000$, and agree very well with the analytical
calculations.
}\label{fig1}
\end{figure}
%%%%%%%%%%%%%%%%%%%%%%%%%%%%%%%%%%%%%%%%%%%%%%%%%%%%%%%%%%%%%%%%%%%%%%%

Fig.~\ref{fig1}(a) shows the $\epsilon$-dependence of the HBT radius 
parameter (\ref{eq20}). For fixed bin width $\epsilon$,
the approximation of the true HBT radius parameter by 
$R_{\rm class}^2(\epsilon)$ is seen to become better with 
increasing $P_0$, in agreement with (\ref{eq22}).
For the HBT radius obtained from the ``quantum'' version of the
Zajc model and depicted in Fig.~\ref{fig1}(b), the situation is 
both qualitatively and quantitatively different. Now, the
HBT radius is always positive, since
the Gaussian wave packets take quantum mechanical localization
properties automatically into account. Also, the $s$-dependence
of the radius is seen to be much weaker since the wave packets
smear out the unphy\-si\-cally strong position-momentum correlations
present in the classical distribution $\rho_{\rm class}^{\rm Zajc}$. 
The HBT radius depends not only on the geometrical size $R_0$, and 
on the momentum width $P_0$ of the source, but also on the wave packet 
width $\sigma$. As seen in Fig.~\ref{fig1}(b), this wave packet
width affects the HBT radius and its $P_0$-dependence significantly 
for $\sigma > R_0$.

In Figs.~\ref{fig1}(c,d) we present the two-particle
correlation functions corresponding to these HBT radius parameters
for characteristic values of the model parameters. The analytical
results, obtained by plotting (\ref{eq19}) and (\ref{eq24}), are
compared to the results from the event generator algorithms (\ref{eq6}) 
and (\ref{eq13}). The plot was obtained using $50$ events of multiplicity 
$1000$. The differences between analytical and numerical results 
originate from statistical fluctuations and become smaller with 
increasing number of events $N_{\rm ev}$ or event multiplicity $N_m$. 
The good agreement between analytical and numerical results in 
Figs.~\ref{fig1}(c,d) indicates the relatively low statistical 
requirements of our algorithms. The reason is that both algorithms 
associate with the {\it discrete} set of generated momenta $\check{\bp}_i$ 
{\it continuous} functions of the measured momenta $\bp_1$, $\bp_2$.
This smoothens any statistical fluctuation significantly.
For the ``classical'' algorithm, small deviations between numerical
and analytical results are still visible in Fig.~\ref{fig1}(c),
while the results of the ``quantum'' algorithm coincide
within line width. This can be traced back to the Gaussian
prefactor $\exp\left( -\sigma^2\, \bq^2/2\right)$ in (\ref{eq6}) 
which provides an additional smoothening of statistical fluctuations 
not present in the ``classical'' algorithm.

%%%%%%%%%%%%%%%%%%%%%%%%%%%%%%%%%%%%%%%%%%%%%%%%%%%%%%%%%%%%%%%%%%%%%%%%%%
\section{Two-Particle Correlations from VNI}
\label{sec4}
%%%%%%%%%%%%%%%%%%%%%%%%%%%%%%%%%%%%%%%%%%%%%%%%%%%%%%%%%%%%%%%%%%%%%%%%%%

\begin{figure}[ht]
\begin{minipage}{6truecm}
\epsfxsize=8truecm
\epsfbox{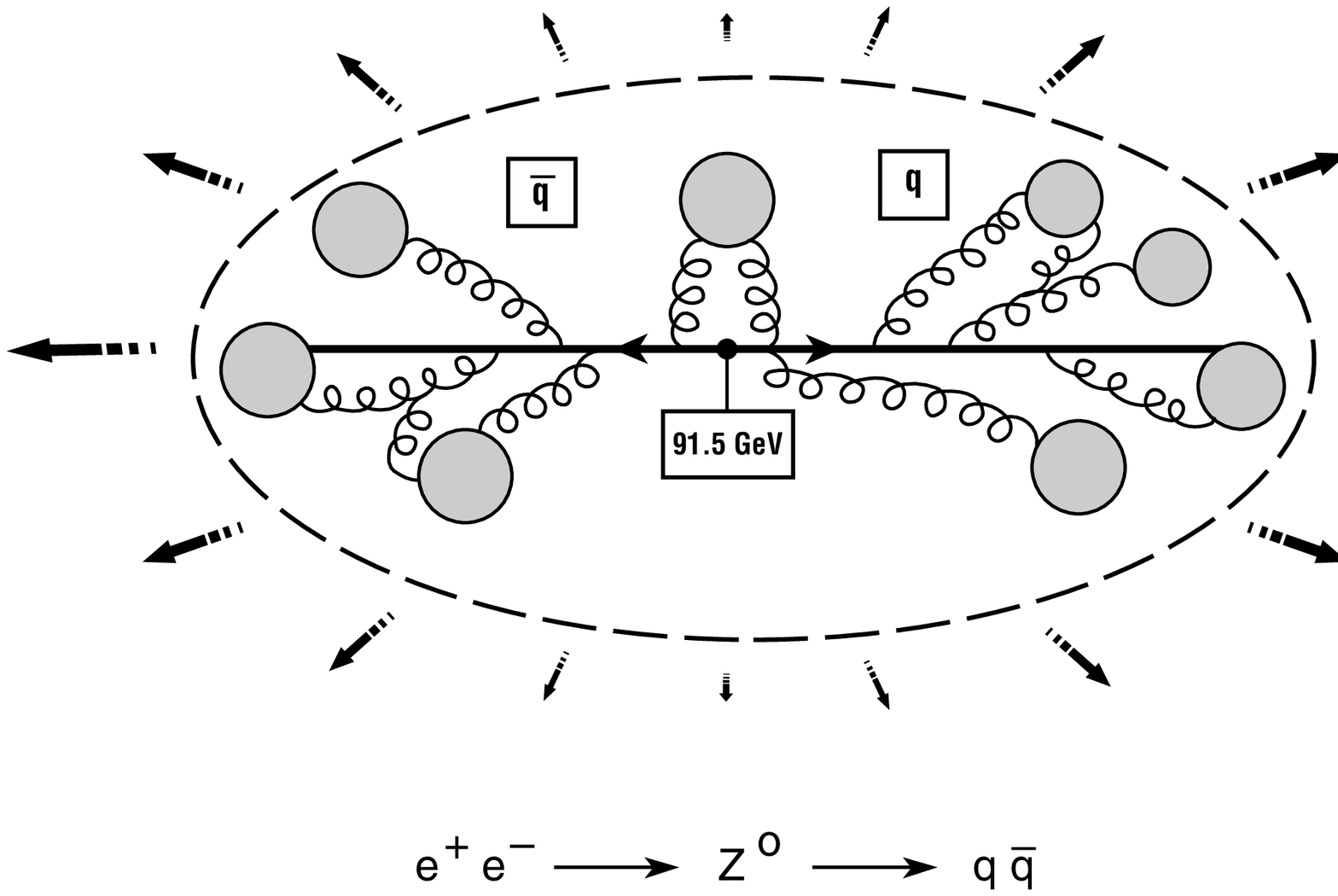}
\end{minipage}
\begin{minipage}{6truecm}
\epsfxsize=8truecm
\hfill{\epsfbox{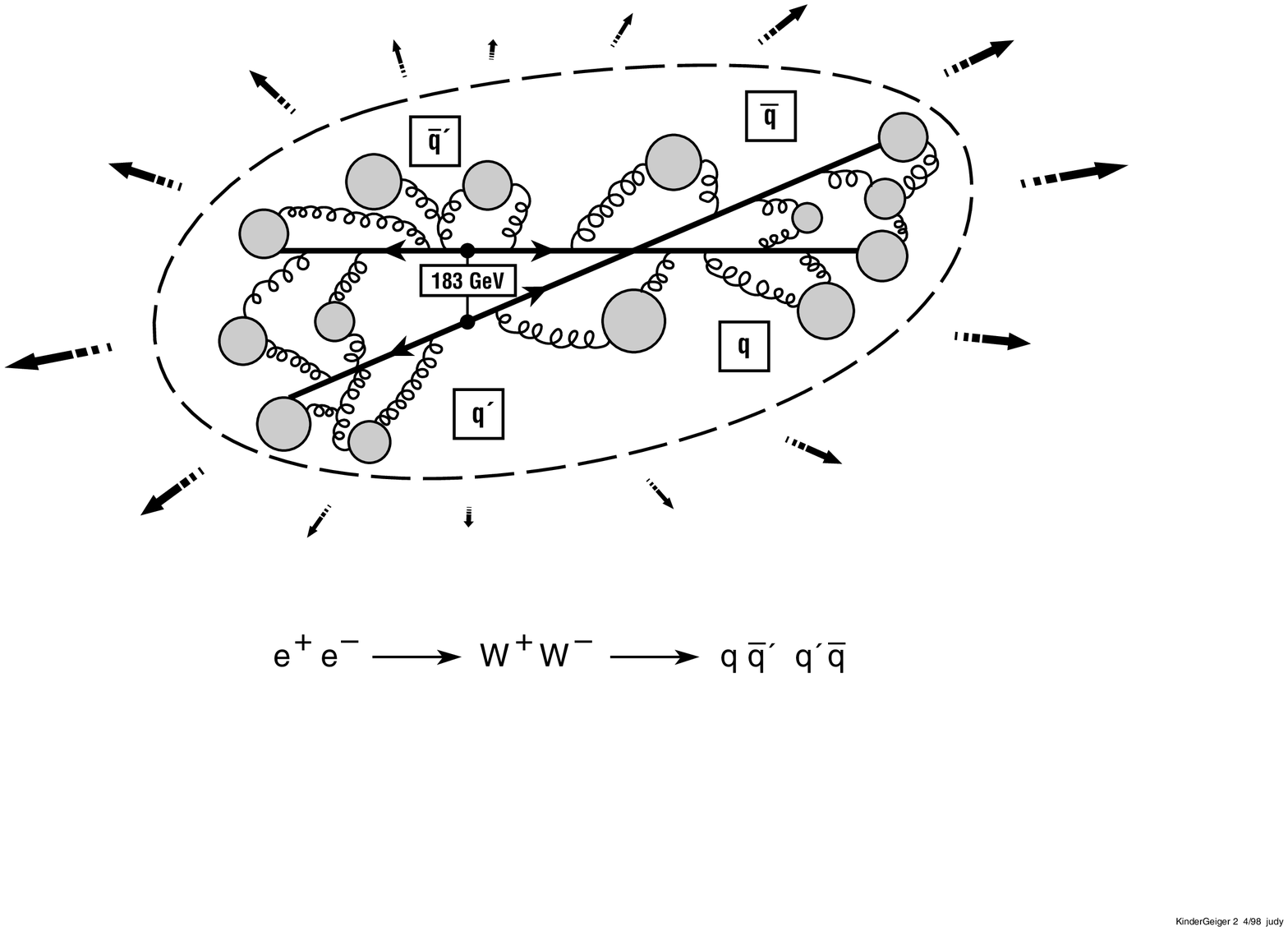}}
\end{minipage}
\caption{
Schematics of the two $e^+e^-$ event types (\ref{Z0}) and (\ref{WW}):
The final-state hadron distribution in $Z^0$ events (left) is due to 
exclusively `endogamous' hadronization of the partonic offspring from 
the $q\bar{q}$ dijet, whereas in $W^+W^-$ events (right) there is, in 
addition, the possibility of `exogamous' hadron production involving a 
mating of partons from the two different $W^+ \rightarrow q\bar{q}'$ and
$W^- \rightarrow q'\bar{q}$ dijets.
\label{fig2}
}
\end{figure}

We have applied the ``quantum'' algorithm discussed above to 
simulated $e^+e^-$ collisions at LEP\,I \cite{LEP1} and LEP\,II \cite{LEP2}
energies from VNI. We have focussed on the channels (see Fig.~\ref{fig2})
 \begin{eqnarray}
    e^+ e^- & \rightarrow &  Z^0 \; \rightarrow q\bar{q} \;\rightarrow \;
   {\rm hadrons} \quad \mbox{at $\sqrt{s} = 91.5$ GeV} \, ,
 \label{Z0}
 \\
    e^+ e^- & \rightarrow &  W^+W^- \; \rightarrow q\bar{q}' \,q'\bar{q} \;
    \rightarrow \;{\rm hadrons} \quad \mbox{at $\sqrt{s} = 183$ GeV} \, ,
 \label{WW}
 \end{eqnarray}
%
%%%%%%%%%%%%%%%%%%%%%%%%%%%% Fig. 2 %%%%%%%%%%%%%%%%%%%%%%%%%%%%%%%%%
%%%%%%%%%%%%%%%%%%%%%%%%%%%%%%%%%%%%%%%%%%%%%%%%%%%%%%%%%%%%%%%%%%%%%%%
%
which provide the ``cleanest'' environment for the study of 
Bose-Einstein correlations in high-energy particle collisions.
Especially for the first channel there exists an impressively 
extensive and accurate data sample of several million events. 
For high-energy $e^+e^-$ collisions, theoretical studies of Bose-Einstein 
enhancements have mainly been performed within the context of the 
string models \cite{strings}, which have been quite successful in 
reproducing the distributions of identical particle pairs on the level 
of 1-parameter Gaussian fits of the correlator \cite{BEtheory}. 
Our study~\cite{GEHW98} is based on the event generator VNI and aimed
at a full inclusion of the space-time structure of the events and
a multidimensional correlation analysis.

The ``classical'' and ``quantum'' interpretation of the event
generator output discussed in section~\ref{sec2b} and \ref{sec2a}
provide {\it two different prescriptions} for the calculation of  
two-particle correlations. This ambiguity in the calculational
scheme can be traced back to the fact that both algorithms 
are {\it a posteriori} remedies for an incomplete quantum
mechanical time evolution which does not account properly for
the quantum mechanical symmetrization of identical $N$-particle
states. The ambiguity has to be removed
by a physical consistency requirement. From our discussion of
the Zajc model in section~\ref{sec3}, first crude statements about 
such consistency requirements can be made: 
\begin{enumerate}
\item
The ``quantum'' interpretation introduces a wave packet width $\sigma$
which must be adjusted to data. The measured HBT radius parameters provide 
an upper bound on $\sigma$, $\sigma < R_{\rm HBT}$. Hence, the 
``quantum'' interpretation can only be consistent with 
experimental data if the wave packet width is not too large.   
\item
In order for the ``classical'' interpretation of the event generator
output to be consistent with experimental data, the generated output
must not be peaked too strongly in phase-space. Otherwise, the
HBT radius may be unphysically small or even show the wrong sign,
cf. section~\ref{sec3a}.
\end{enumerate}
The first step in a realistic study of two-particle correlations 
is necessarily to refine these crude statements. To what extent do 
physical observables depend on the choice of algorithm? Which choice 
of $\sigma$ is consistent with experiment? Are there phenomenological 
reasons to prefer one of the algorithms? These workshop proceedings 
reflect the state of our work when Klaus left us. At the time of the 
airline accident we had only completed the calculation with the 
``quantum'' algorithm, and only for a single value of the wave packet 
width $\sigma$. A comparison of the two algorithms remains to be done.

%%%%%%%%%%%%%%%%%%%%%%%%%%%%%%%%%%%%%%%%%%%%%%%%%%%%%%%%%%%%%%%%%%%%%%%%%
\subsection{Two-particle correlations at vanishing pair momentum}
\label{sec4a}
%%%%%%%%%%%%%%%%%%%%%%%%%%%%%%%%%%%%%%%%%%%%%%%%%%%%%%%%%%%%%%%%%%%%%%%%%

\begin{figure}[ht]
\vspace*{1cm}
\epsfxsize=7.2cm 
\begin{minipage}{70mm}
\centerline{ \epsfbox{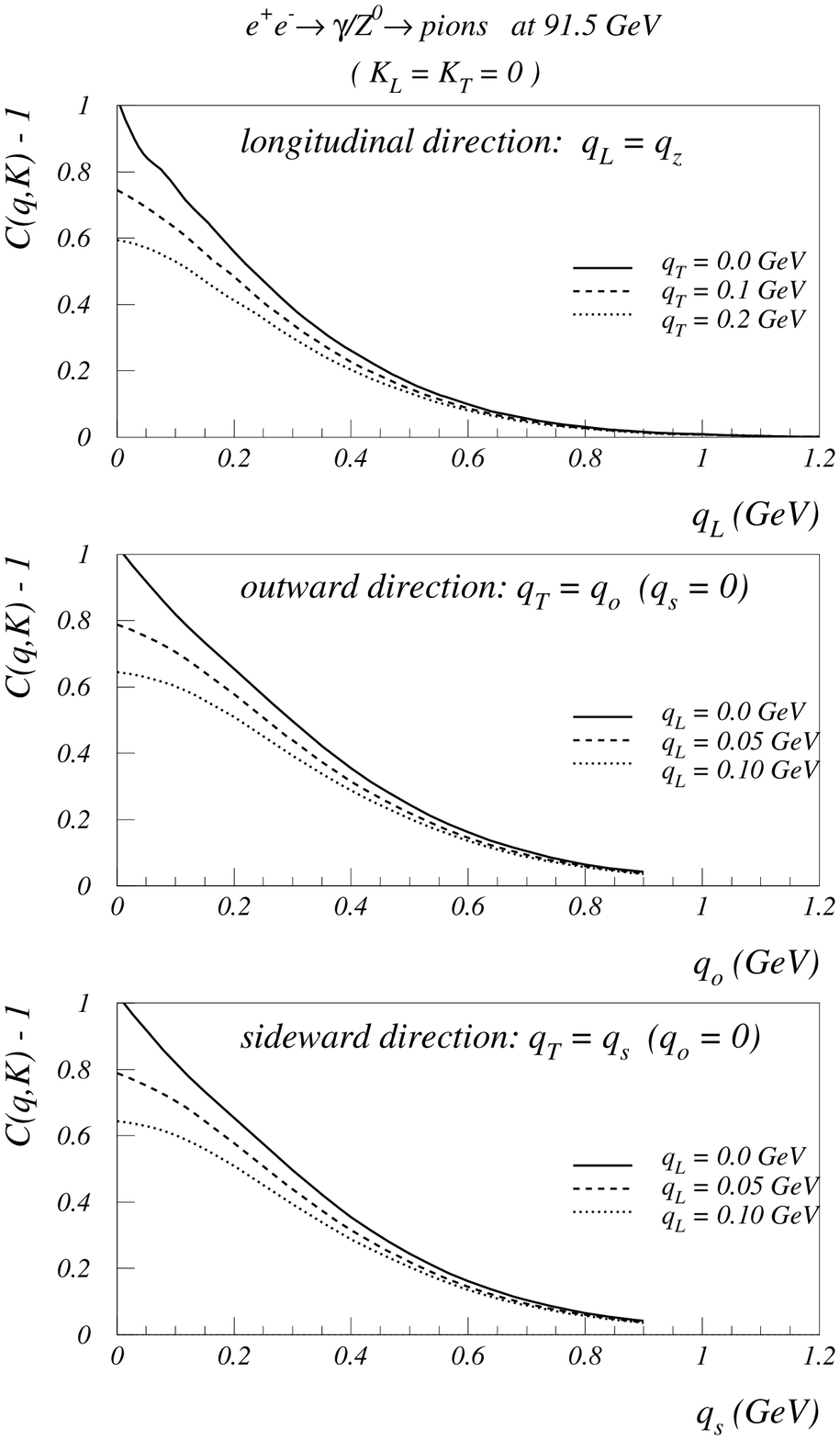} }
\end{minipage}
%%%%%%%%%%%%%%%%%%%%%%%%%%%%%%%%%%%%%%%%%%%%%%%%%%%%%%%%%%%%
\epsfxsize=7.2cm
\begin{minipage}{70mm}
\centerline{ \epsfbox{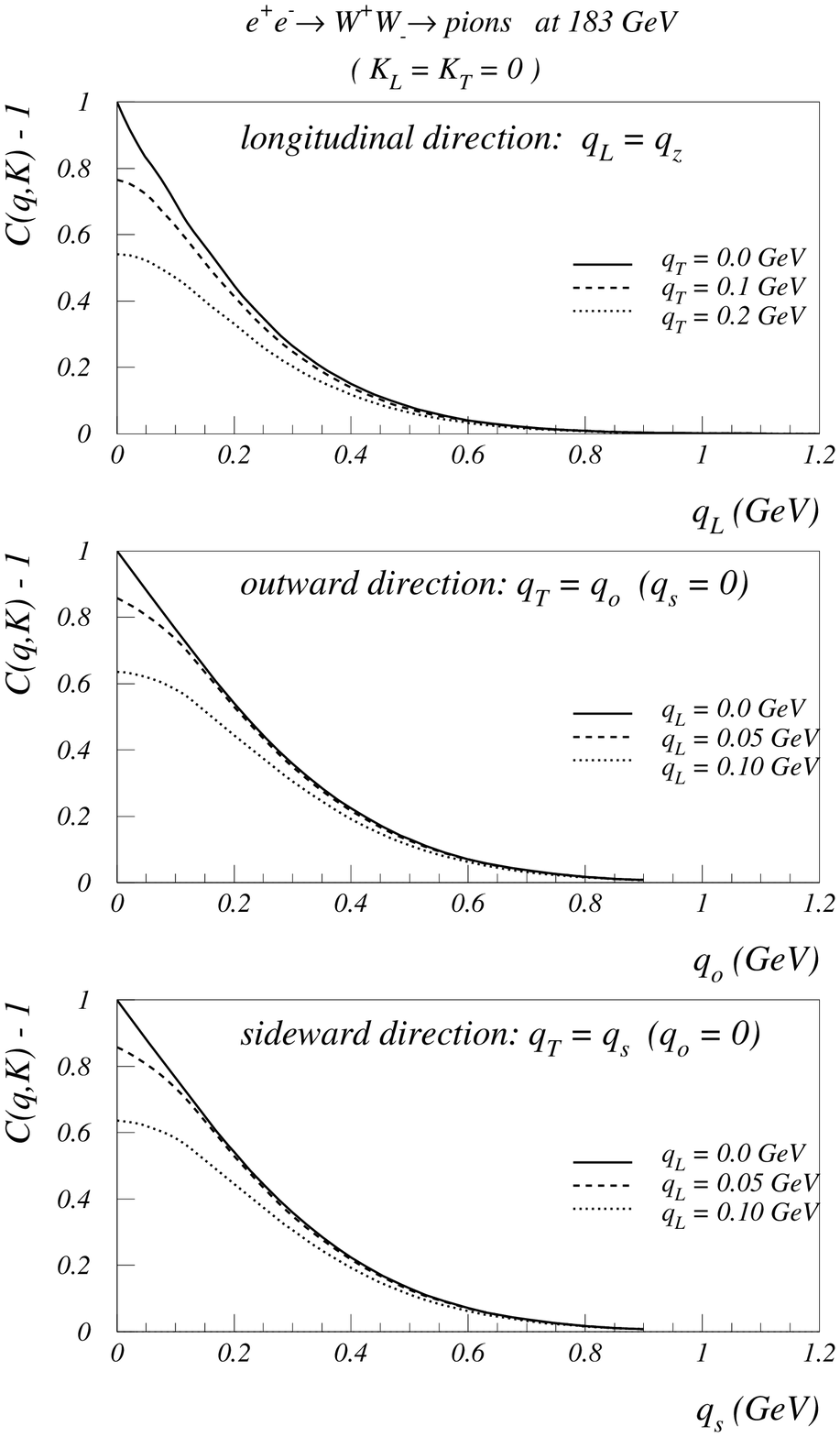} }
\end{minipage}
\caption{
The correlation function of same-sign pions for different values of 
the relative pair momentum $\bq$ for vanishing pair momentum $\bK$,
$C(q_z,q_s,q_o,\bbox{0})-1$.}
\label{fig3}
\end{figure}

Fig.~\ref{fig3} shows the correlator $C(\bq,\bK) - 1$ for different 
$\bq$-values and vanishing pair momentum $\bK$ in the c.m. frame of 
the collision, $C(q_z,q_s,q_o,\bbox{0}) - 1$. The widths of the 
correlator in three Cartesian directions are roughly 
the same. Quantitatively, they are roughly given by the inverse of
the wave packet width $\sigma$ which suggests that they are dominated 
%
%%%%%%%%%%%%%%%%%%%%%%%%%%%%%%%%%%%%%%%%%%%%%%%%%%%%%%%%%%%%
%%%%%%%%%%%%%%%%%%%%%%%%%%%%%%%%%%%%%%%%%%%%%%%%%%%%%%%%%%%%%%%%%%%%%
%
by the ``quantum mechanical smearing'' features of the ``quantum'' 
Bose-Einstein algorithm used here. A simulation with the ``classical''
algorithm remains to be done.

\begin{figure}[ht]\epsfxsize=13.5cm 
\centerline{\epsfbox{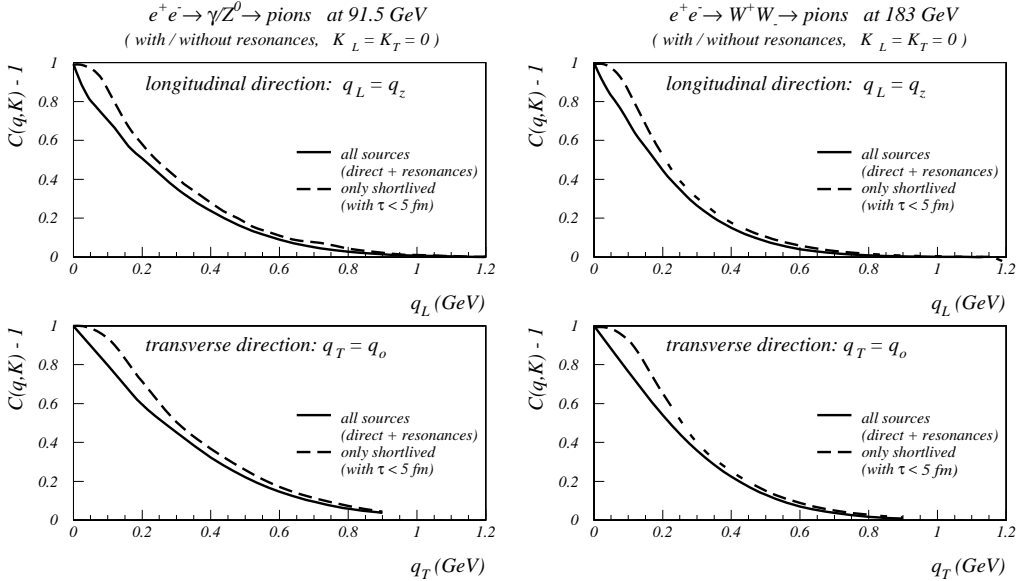}}
\caption{ The correlator $C(\bq,\bK) - 1$ for $\bK = 0$
with (solid lines) and without (dashed lines) the contributions of
pions stemming from long-lived resonances.
}\label{fig4}
\end{figure}

One also sees that near $\bq=0$ the correlation functions show 
characteristic deviations from a Gaussian shape. However, at non-zero 
values of the orthogonal $\bq$-components, the correlators become 
nicely Gaussian. The non-Gaussian features at small $\bq$ can be traced 
to decay contributions from long-lived resonances: Fig.~\ref{fig4} 
shows the same correlator 
%
%%%%%%%%%%%%%%%%%%%%%%%%%%%%%%%%%%%%%%%%%%%%%%%%%%%%%%%%%%%%%%%%%%%%
%%%%%%%%%%%%%%%%%%%%%%%%%%%%%%%%%%%%%%%%%%%%%%%%%%%%%%%%%%%%%%%%%%%%%%%
%
with and without such contributions. Neglecting the pions from 
long-lived resonance decays, the correlation function becomes Gaussian
and wider, reflecting a smaller source size. With long-lived resonances 
included, the effective pion source $S(x,\bK)$ is larger, resulting in 
a narrower and non-Gaussian correlation function. This clearly 
demonstrates the sensitivity of the correlation function on the 
space-time {\em geometry} of the source function $S(x,\bK)$.

%%%%%%%%%%%%%%%%%%%%%%%%%%%%%%%%%%%%%%%%%%%%%%%%%%%%%%%%%%%%%%%%%%%%%
\subsection{Pair momentum dependence of the correlation function}
\label{sec4b}
%%%%%%%%%%%%%%%%%%%%%%%%%%%%%%%%%%%%%%%%%%%%%%%%%%%%%%%%%%%%%%%%%%%%%%%

\begin{figure}[ht]
\vspace*{1cm}
\epsfxsize=7.4cm
\begin{minipage}{70mm}
\centerline{ \epsfbox{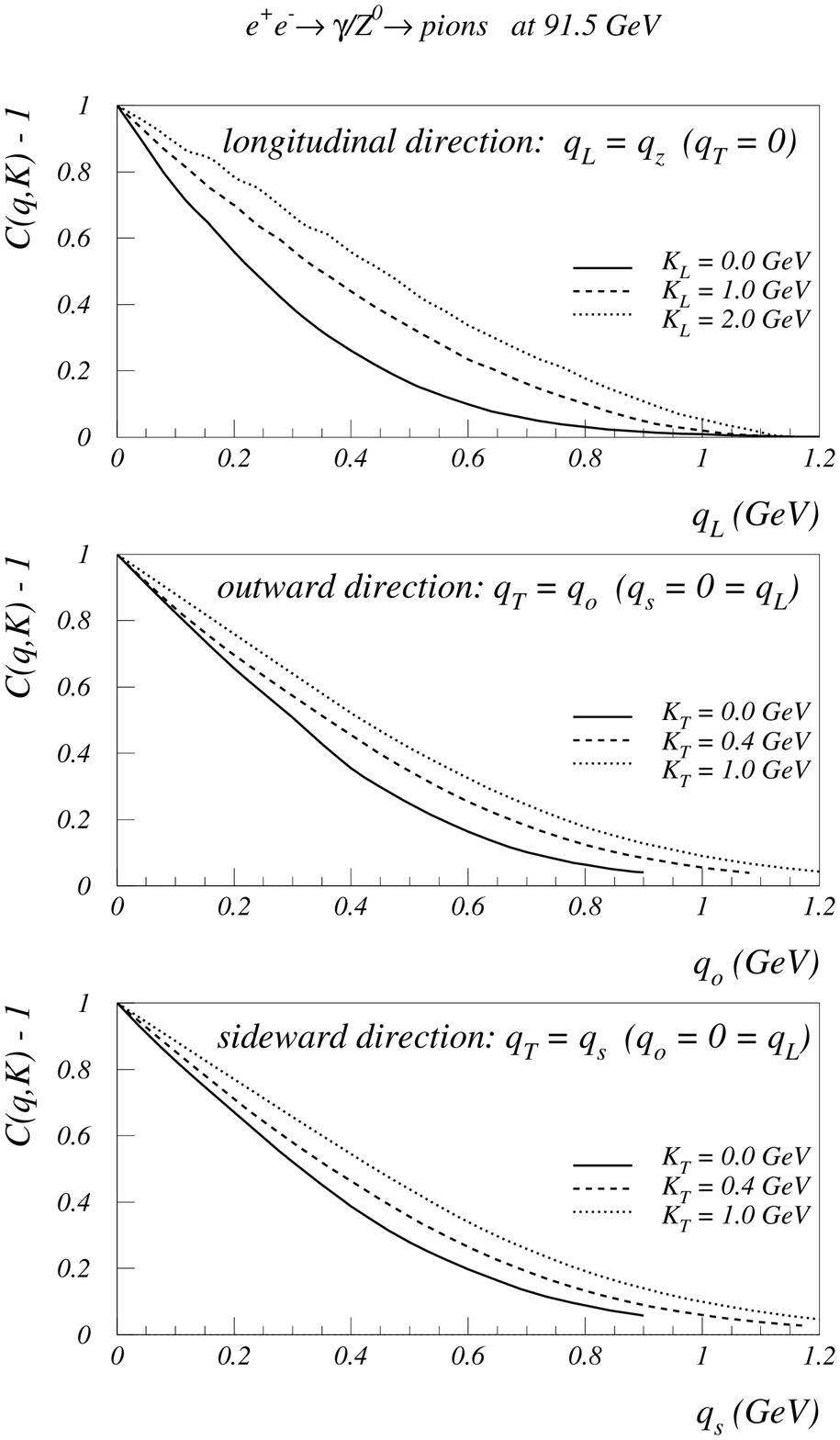} }
\end{minipage}
%%%%%%%%%%%%%%%%%%%%%%%%%%%%%%%%%%%%%%%%%%%%%%%%%%%%%%%%%%%%
\epsfxsize=7.4cm
\begin{minipage}{70mm}
\centerline{ \epsfbox{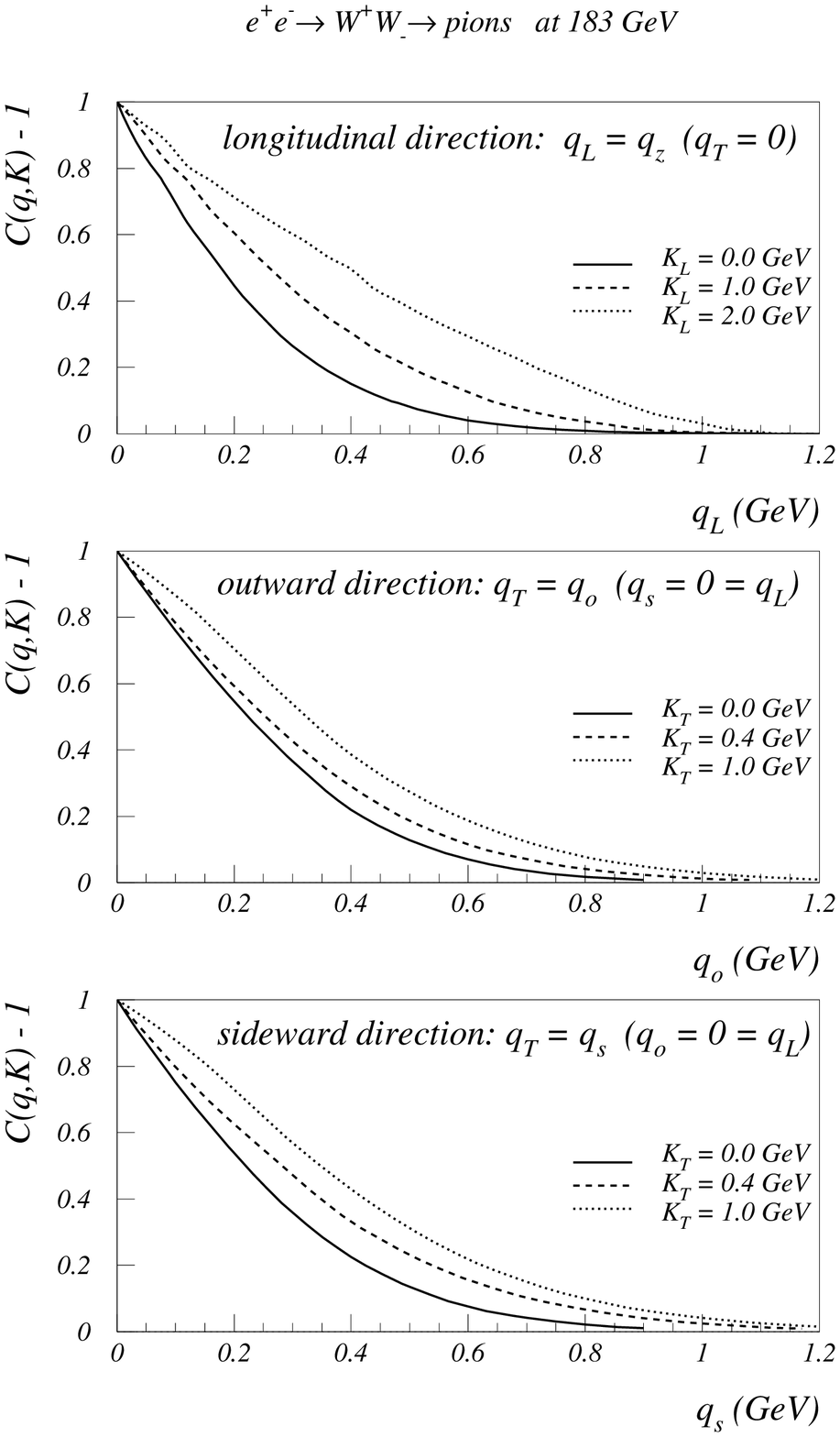} }
\end{minipage}
\caption{
The correlation function of same-sign pions $C(\bq,\bK)-1$ for various
values of the pair momentum $\bK = (K_L,\bK_\perp)$, where $K_L = K_z$ 
is the direction along the thrust axis and $K_\perp = |\bK_\perp|= 
\sqrt{K_x^2+K_y^2}$ is the momentum transverse to it. The correlators
are plotted against one component of the relative momentum, setting the 
two other components to zero.}
\label{fig5}
\end{figure}

As discussed before, the dependence of the correlation function
on the pair momentum $\bK$ indicates $x$-$K$-correlations in the source 
and can thus be sensitive to the space-time {\em dynamics} of hadron 
production. Fig.~\ref{fig5} shows the correlation function 
$C(\bq,\bK) -1$ of same-sign pions for various values of the pair 
momentum $\bK = (K_L,\bK_\perp)$. Clearly, as $\bK$ increases, the 
correlation function becomes wider, indicating a smaller effective 
source, qualitatively (although not quantitatively) very similar to the
corresponding observations in heavy-ion collisions. This is an 
interesting prediction which to our knowledge has not yet been tested 
in the LEP experiments. According to the more detailed analysis 
presented in \cite{GEHW98} the observed $\bK$-dependence can be fully 
attributed to the $\bK$-dependence of resonance decay contributions; 
however, this may be to some extent an artifact of the employed 
``quantum'' Bose-Einstein algorithm whose wave packet width $\sigma$ 
apparently dominates the widths of the correlation functions shown above. 

%%%%%%%%%%%%%%%%%%%%%%%%%%%%%%%%%%%%%%%%%%%%%%%%%%%%%%%%%%%%%%%%%%%%%%%%%
\section{Outlook}
\label{sec5}
%%%%%%%%%%%%%%%%%%%%%%%%%%%%%%%%%%%%%%%%%%%%%%%%%%%%%%%%%%%%%%%%%%%%%%%%%

The rather abrupt ending of the above section on physics results
gives a sad feeling for the gap left by Klaus. In our collaboration,
he was the only one who actually knew how to run VNI. Our original 
%
%%%%%%%%%%%%%%%%%%%%%%%%%%%%%%%%%%%%%%%%%%%%%%%%%%%%%%%%%%%%
%%%%%%%%%%%%%%%%%%%%%%%%%%%%%%%%%%%%%%%%%%%%%%%%%%%%%%%%%%%%%%%%%%%%%
%
motivation came from relativistic heavy-ion collisions where the 
space-time geometry and dynamics of the event plays a crucial role 
in understanding basic measurable quantities. Mainly for this reason 
we developed algorithms which allow to calculate identical 
two-particle correlation functions from an arbitrary event generator 
output. We tested the accuracy and statistical requirements of these 
algorithms in simple toy model studies and applied them in a first 
realistic study to the hadronic channels in $e^+$ $e^-$
annihilation at the $Z^0$-peak and near the $W^+$-$W^-$ threshold. 
However, a comparative study of both algorithms is still missing and our 
main goal, the application of these algorithms to the study of 
event simulations of heavy ion collisions, is not achieved yet. We 
plan to do so in the near future. Also, we plan to make the algorithms 
described in section~\ref{sec2} publicly available, using the 
Open Standard for Codes and Routines (OSCAR) format, advocated 
in~\cite{YP97}. 

%%%%%%%%%%%%%%%%%%%%%%%%% References %%%%%%%%%%%%%%%%%%%%%%%%%%%%%

\end{document}